\documentclass[conference]{IEEEtran}

\usepackage{caption} 
\usepackage{amsmath}
\usepackage{pbox}
\usepackage{url}

\usepackage{graphicx}
\usepackage{bbold}
\usepackage[outdir=./]{epstopdf}
\begin{document}
\title{Probabilistic Matching: Causal Inference under Measurement Errors}

\author{\IEEEauthorblockN{Fani Tsapeli}
\IEEEauthorblockA{School of Computer Science\\
University of Birmingham\\
Birmingham, UK\\
Email: t.tsapeli@cs.bham.ac.uk}
\and
\IEEEauthorblockN{Peter Tino}
\IEEEauthorblockA{School of Computer Science\\
University of Birmingham\\
Birmingham, UK\\
Email: P.Tino@cs.bham.ac.uk}
\and
\IEEEauthorblockN{Mirco Musolesi}
\IEEEauthorblockA{Department of Geography\\
University College London\\
London, UK\\
Email: m.musolesi@ucl.ac.uk}
}

\maketitle

\begin{abstract}
The abundance of data produced daily from large variety of sources has boosted the need of novel approaches on causal inference analysis from observational data. Observational data often contain noisy or missing entries. Moreover, causal inference studies may require unobserved high-level information which needs to be inferred from other observed attributes. In such cases, inaccuracies of the applied inference methods will result in noisy outputs. In this study, we propose a novel approach for causal inference when one or more key variables are noisy. Our method utilizes the knowledge about the uncertainty of the real values of key variables in order to reduce the bias induced by noisy measurements. We evaluate our approach in comparison with existing methods both on simulated and real scenarios and we demonstrate that our method reduces the bias and avoids false causal inference conclusions in most cases.   
\end{abstract}

\section{Introduction}
Nowadays, there is an increasing data availability. Smartphones and wearables sensors, social media, web browsing information and sales recordings are only few of the newly available information sources. Discovering dependencies and patterns within such datasets could provide useful insights to businesses or social scientists studying human behavior. 

An important part of such studies involves the discovery of correlation or causation links among factors of interest. For example, several studies examine the impact of sentiment and opinions expressed through social media to traded assets prices \cite{bollen2011twitter, zhang2011predicting, tsapeli2016model}. Others study the impact of location, activity and communication patters on people mood \cite{lathia2013smartphones}, stress level \cite{tsapeli2015investigating} or eating and sleeping disorders \cite{tsapeli2015investigating} by processing smartphones sensor data. In such studies, the variables of interest are usually not directly measured. Instead, they are inferred from raw data. For example, in \cite{bollen2011twitter, tsapeli2016model} a sentiment index is inferred from Twitter data by applying text processing and classification techniques. However, such procedures result in inaccurate estimation of the variables of interest. Moreover, several studies have shown that social media data in some countries have undergone censorship \cite{bamman2012censorship}. In such cases, sentiment or opinion tracking could be biased. Also, in \cite{tsapeli2015investigating}, location context (e.g. home, work, entertainment place etc.) is used to understand the causal link between location and stress level. However, the real location context is not known; instead it is inferred from smartphone sensors and consequently it could be inaccurate. 

Causal discovery when key variables are unobserved or inaccurately measured is a particularly challenging task. Latent variable models have been used to handle such cases \cite{fornell1981structural, bentler1980multivariate}. Such models include one or more unobserved or latent variables. 
% In a broad sense, every regression model that includes an error term is a latent variable model, since the error term represents the effect of unmeasured factors.
Scientists usually attempt to estimate the values of a latent variable from other observed variables by fitting the data in a structural equation model \cite{fornell1981structural}. However, the selection of a proper model is a complex task which may result in misspecification and overfitting. 

Matching design \cite{william2002experimental} is an alternative causal inference approach which does not require fitting data to a structural equation model. The main idea is that the average impact of a treatment variable (or predictor) to an outcome or effect variable can be approximated by comparing the outcome variable values of units with similar characteristics and different treatment values. In order to minimize any bias due to differences on units characteristics (\textit{confounding bias}), matching methods attempt to find optimal pairs of units. When the values of these characteristics are inaccurately measured or inferred from other variables, matching may fail to sufficiently eliminate any confounding bias which may result in misleading conclusions about the causal relationship of the examined variables. 

In this work, we propose \emph{probabilistic matching}, a matching method that takes into account the uncertainty about the real values of a noisy variable and attempts to find optimal pairs of units in order to maximize the probability that the matched units have similar characteristics. Our method is based on the assumption that a probability distribution describing the real values of each unobserved variable is known or can be approximated. Although this assumption may restrict the applicability of the proposed method, it is realistic in many scenarios. For example, when an inference procedure is applied in order to learn the values of an unobserved variable $L$ from some observed attributes $C$, a probability distribution $Pr(L|C)$ can be approximated, as we discuss later in Section \ref{sec:evaluation}. 
  
We evaluate the proposed matching framework on two different simulation studies in comparison with a conventional matching method. We demonstrate that probabilistic matching reduces significantly the confounding bias and results in more accurate causal conclusions. We also evaluate our method on a real dataset. In particular, we use the social media dataset described in \cite{MicroBlogPCU} in order to test whether text messages containing URLs tend to be reposted more often. This dataset includes a rich variety of features extracted from Weibo microblogging service for 111 users along with a manually assigned binary label for each user indicating whether he/she has been characterized as \emph{spammer} or not. In our scenario, we assume that the \emph{spammer} label is an unobserved confounding variable and we apply a spammer detection method \cite{benevenuto2010detecting} in order to infer a label for each user from other observed attributes. We map the classification outputs to probability distributions describing the probability of a user to be a \emph{spammer} and we use these probability distributions to our matching framework. We demonstrate that the number of URLs in text messages indeed influences the number of reposts. We repeat our causality study by applying a conventional matching method in two scenarios: 1) the ground truth binary \emph{spammer} identifier is known and 2) only the noisy \emph{spammer} identifier inferred from the data is known. The results of the first scenario serve as the ground-truth. We demonstrate that our results come in agreement with the conclusions of the first scenario, while the examined conventional matching method fails to detect the causal link.

\section{Background}

In this section we provide some background knowledge on causal inference based on the matching design framework. For more clarity, we summarize the notation used in the paper in Table \ref{tab:notation}. According to Rubin's framework \cite{rubin2011causal}, the causal impact of a binary treatment  on a unit $u$ can be assessed by comparing the outcome $Y_1(u)$, if the unit has received the treatment, with the outcome $Y_0(u)$, if the unit has not received the treatment. The \emph{fundamental problem of causal inference} is that it is not possible to observe both $Y_1(u)$ and $Y_0(u)$ for the same unit $u$. If treatment is randomly assigned, the average treatment effect (ATE) can be estimated as $E\{Y_1\} - E\{Y_0\}$, where $E\{Y_1\}$ and $E\{Y_0\}$ are expectations w.r.t. uniform distribution over treated and untreated units, respectively. However, if the treatment assignment is conditional on certain characteristics of the units, the causal inference could be biased. The  characteristics of units are called \emph{confounding variables}. Matching methods attempt to eliminate this bias by comparing the outcome values of units with similar observed characteristics. In particular, if $U$ is a set of treated units and $V$ is a set of control units (i.e. units which have not received the treatment), matching methods match each treated unit $u \in U$ with the \emph{"most similar"} control unit $v \in V$. If $G$ is the set of matched pairs of units, the average treatment effect is estimated as $E_{(u,v)\in G}\{Y_1(u) - Y_0(v)\}$, where the expectation is with respect to uniform probability distribution over $G$. The (dis)similarity between units is measured as a distance between their confounding variable values (for some metric). 

\begin{table}[]
\begin{small}
\centering
\begin{tabular}{|l|l|}
	\hline
 	\textbf{Symbol} & \textbf{Description}\\
  \hline
 $N$ & Number of units\\
  \hline
 $P$ & Number of confounding variables\\
  \hline   
 $Y$ & Outcome variable, described with a $1 \times N$ vector\\
  \hline
  $y_u$ & The outcome value of unit $u$\\
   \hline
  $X$ & Treatment variable, described with a $1 \times N$ vector\\
 \hline
 $x_u$ & The treatment value of unit $u$\\
  \hline
  $Z$ & $P \times N$ matrix of confounding variables \\
 \hline
  $z_u$ & \pbox{10cm}{the $u^{th}$ column of $Z$, denoting a $P \times 1$ vector of\\ values of unit $u$ for the $P$ confounders} \\
 \hline
  $z^p$ & \pbox{10cm}{the $p^{th}$ line of $Z$, denoting a $1 \times N$ vector of\\ values of the $N$ units for the $p^{th}$ confounder} \\
 \hline
  $z^p_u$ & \pbox{10cm}{element in column $u$ and line $p$ of $Z$, denoting \\the value of unit $u$ for the $p$ confounder} \\ 
 \hline
  $G$ & Set of matched treated and control units\\
 \hline
  $G_U$ & Set of matched treated units\\
 \hline
  $G_V$ & Set of matched control units\\
 \hline
  $\tilde{L}$ & \pbox{10cm}{Variable with measurement errors, described \\with a $1 \times N$ vector}\\
 \hline
 $L_u$ & Random variable with $Pr(L_u|\tilde{L} = \tilde{l}_u)$\\
 \hline
 $X_u$ & \pbox{10cm}{stochastic variable describing the treatment of $u$}\\
 \hline
 $\mathbf{X}$ & \pbox{10cm}{$1 \times N$ vector of stochastic variables describing the \\treatments of the $N$ units}\\
 \hline
 $\mathbf{Z}$ & \pbox{10cm}{$P \times N$ matrix of stochastic confounders}\\
 \hline
  $Z_u$ & \pbox{10cm}{$u^{th}$ column of $\mathbf{Z}$, denoting a $P \times 1$ vector of\\ random variables for unit $u$} \\
 \hline
  $Z^p$ & \pbox{10cm}{$p^{th}$ line of $\mathbf{Z}$, denoting a $1 \times N$ vector of\\ random variables for the $p$ confounder} \\
 \hline
  $Z^p_u$ & \pbox{10cm}{element in column $u$ and line $p$ of $\mathbf{Z}$, denoting a\\ random variable for the $p^{th}$ confounder of unit $u$} \\
 \hline
 $\mathcal{D}(z_u, z_v)$ & Distance between vectors $z_u$, $z_v$\\
 \hline
 $D(Z_u^p, Z_v^p)$ & Distance between random variables $Z_u^p$, $Z_v^p$\\
 \hline
 $\mathbf{D}(Z_u, Z_v)$ & \pbox{10cm}{Distance between random variables  vectors $Z_u$, $Z_v$} \\
 \hline
 $\mathbb{D}_{Z_u, Z_v}$ & \pbox{10cm}{$P \times 1$ vector of distances between the $P$ \\random variables $Z_u^p$, $Z_v^p$} \\
 \hline
 $\Delta(u, v)$ & Distance between units $u$, $v$\\
 \hline
\end{tabular}
\caption{Notation.}
\label{tab:notation}
\end{small}
\end{table}

\subsection{Matching Methods}
\label{sec:matchingMethods}
Several methods for creating pairs of units $(u,v)\in G$ have been proposed. The matching methods involve two steps: 1) the matching and 2) the balance check. In the matching step, a method that creates pairs of treated and control units $(u,v)\in G$ based on closeness of their confounding variables is applied using some notion of a distance between confounding variable values. Afterwards, in the balance check step, the remaining confounding bias due to imperfectly matched units needs to be estimated. If the resulted groups of matched treated and control units are not adequately balanced, the matching method needs to be revised. The balance can be examined by checking the standardized mean difference between the treated and control units, by applying a t-test or a Kolmogorov-Smirnov test, or by examining the quantiles of the matched units. This checking has to be done for each confounding variable.  

% Several methods have been used for matching. The most straightforward method is \textit{Nearest Neighbor Matching}, which matches each treated unit to the control unit with the lowest distance on the corresponding confounding variable values. Other methods, which attempt to optimize a global distance measure among all matched pairs have been proposed \cite{}. In this work we'll focus on \emph{Genetic Matching}, a popular matching method which uses a generalized weighted Mahalanobis distance and applies an evolutionary search algorithm to determine the weight of each confounder.

The simplest matching method is the \emph{Nearest Neighbor Matching} which matches each treated unit to the control unit with the lowest distance between the corresponding confounding variable values. Another popular matching method is \emph{genetic matching} \cite{diamond2013genetic}.
Let us denote by $N$ the number of units in the study, $P$ the number of confounding variables and $Z$ a $P \times N$ matrix of confounding variables with $z_u$ the $u^{th}$ column of $Z$, i.e. $P \times 1$ column vector of values of unit $u$ for the $P$ confounders. Genetic matching uses as distance metric between confounder vectors the following weighted Mahalanobis distance:
% \cdot (S^{-\frac{1}{2}})^T \cdot W \cdot S^{-\frac{1}{2}}
 \begin{equation}
 	d_{u, v, W} = \sqrt{(z_u - z_v)^T \cdot \mathbf{W} \cdot (z_u - z_v)}
	\label{eq:genMatch}
 \end{equation}
\noindent
where $\mathbf{W} = (S^{-\frac{1}{2}})^T \cdot W \cdot S^{-\frac{1}{2}}$, with W a $P \times P$ diagonal positive definite weight matrix and $S^{-\frac{1}{2}}$ is the Cholesky decomposition of the sample covariance matrix of $Z=[z_1,...,z_N]$. The diagonal elements of $W$ are selected by applying an evolutionary search algorithm that attempts to find the optimal weights  to minimize a loss function. Several loss functions can be used. A commonly used loss is the minimum p-value of a t-test or a Kolmogorov-Smirnov distributional test on the matched pairs of treated and control units resulting from applying a given $W$ in the distance calculations between confounders. The loss is calculated for each confounding variable. Thus, if $p_p$ is the p-value of the $p^{th}$ confounding variable, the objective is to find a matrix $W$ that minimizes the $min_p p_p$. Other loss functions are based on comparisons of the quantiles of confounding variables for the matched treated and control units. In detail, denote by $G_U$ and $G_V$ the sets of matched treated and control units, respectively, i.e. for each pair $(u,v)\in G$, $u \in G_U$ and $v \in G_V$. For $p^{th}$ confounding variable, we think of the corresponding values for matched treated units $\{z^p_u: u \in G_U\}$ as realizations of a random variable $A^p$.
Analogously, the values $\{z^p_v: v \in G_V\}$ of matched control units will be considered realizations of a random variable $B^p$.
Given a set of $K$ quantiles $a^{p}(k)$ and $b^{p}(k)$ of $A^p$ and $B^p$, respectively, we calculate a set of quantile differences $\Delta^p = \{|a^p(k) - b^p(k)|\}_{k=1}^K$. Then, one of the following loss functions can be applied: 1) $mean_p  mean{\Delta}^p$, 2) $max_p {\Delta}^p$, 3) $median_p {\Delta}^p$, 4) $mean_p max \Delta^p$, 5) $max_p max \Delta^p$, 6) $median_p max \Delta^p$, 7) $mean_p median \Delta^p$, 8) $max_p median \Delta^p$ and 9) $median_p median \Delta^p$.

%
% \subsubsection{Matching Parameters}
% \label{sec:MatchingParameters}
% There are three key parameters that can be adjusted when applying a matching method:
% \begin{itemize}
% 	\item \textbf{Many-to-one Matching:} Instead of 1-to-1 matching (i.e. each treated unit is matched with one control unit), k-to-1 matching can be applied (i.e. k controls are used for each treated unit). In this case, the k control units with the lowest distance are selected.
% 	\item \textbf{Matching with Caliper Distance (Threshold):}  A caliper distance can be used to avoid matching units with large distance (i.e. larger than the caliper distance) in cases that a better match cannot be found.
% 	\item \textbf{Matching with Replacement:} Matching with or without replacement is another key consideration in matching. If matching with replacement is applied, each treatment unit can be matched with multiple control units; otherwise, each control unit can be used only once. Matching with replacement can decrease bias when there are multiple treated units with small distance to a single control unit. Since some control units are used multiple times, the analysis could be dependent on the selected control units. Frequency weights need to be used in order to eliminate this bias.
% \end{itemize}

\subsection{Matching with Continuous Treatments}
\label{sec:matchingContinuous}
Although matching frameworks have been proposed mainly for bivariate treatment variables, some recent studies also consider continuous treatments \cite{lu2001matching, hirano2004propensity}. In such cases units cannot be split into treatment and control groups. Instead, each unit can be matched to any other unit. The goal of matching is to create pairs of units with similar values on their confounding variables but different treatment values. In \cite{lu2001matching} the distance between units $u$, $v$ is estimated as follows:

\begin{equation}
	\Delta(u, v) = \frac{{\cal D}(z_u, z_v)+\epsilon}{(x_u-x_v)^2}
	\label{eq:continuousDistance}
\end{equation} 
\noindent
where ${\cal D}(z_u, z_v)$ is the distance between the vectors of confounding variables values of units $u$ and $v$ (this can be the euclidean distance, the Mahalanobis distance or any other distance metric), $\epsilon>0$ a small constant and $x_u$, $x_v$ are the treatment values of $u$ and $v$, respectively. With respect to unit $v$, unit $u$ will be considered as treated if $x_u>x_v$. The average treatment effect is estimated as follows:

\begin{equation}
	E_{(u,v)\in G}\Big\{\frac{y_u - y_v}{x_u - x_v}\Big\}
	\label{eq:avgTrEffect}
\end{equation}

\subsection{Genetic Matching with Continuous Treatments}

To the best of our knowledge, Genetic Matching, so far has been used only for binary treatments. However, it can be extended to continuous treatments by modifying Eq.  \eqref{eq:continuousDistance} as follows:

\begin{equation}
	\Delta(u, v) = \frac{d_{u, v, W}+\epsilon}{|x_u-x_v|}
	\label{eq:genMatchContinuous}
\end{equation} 

The loss function also needs to be modified in order to penalize any matrix $W$ which results in matched units with similar treatments. We think of the absolute differences on the  $p^{th}$ confounding variable values of the matched treated and control units $\{|z^p_u - z^p_v|: (u, v) \in G\}$ as realizations of a random variable $A^p$. Then, we define a set of $K$ quantiles $\Delta^p = \{q^p(k)\}_{k=1}^K$. The loss function can be selected based on this quantiles set as described in Section \ref{sec:matchingMethods}.

\section{Probabilistic Matching}
\label{sec:probMatching}
We further extend the framework for matching with continuous treatments to cases where treatment and/or one or more confounding variables may have noisy or censored measurements. We assume that for each unobserved variable $L$ there is an observed noisy version $\tilde{L}$. For example, $\tilde{L}$ could be a location label inferred from smartphone sensor data (and consequently subject to inaccuracies) and $L$ the real unknown location label. We also assume that for each observation $\tilde{l}_u$ of $\tilde{L}$ the corresponding random variable $L_u$ has known probability distribution $Pr(L_u|\tilde{L} = \tilde{l}_u)$. In the following, we will consider the general case where all the key variables are noisy with the understanding that in the case of no noise the corresponding distribution reduces to the delta function:

\begin{equation}
	Pr(L_u | \tilde{L} = \tilde{l}_u) = \begin{cases}
	1 &\text{, $L_u = \tilde{l}_u$}\\
	0 &\text{, $L_u \neq \tilde{l}_u$}
	\end{cases}
	\label{eq:observedVarPDF}
\end{equation}

Denote by $X_u$ the random variable describing the treatment of unit $u$ and with $\mathbf{X}$ a $1 \times N$ random vector of treatment variables of all units. We also denote by $Z_u^p$ the random variable describing the $p^{th}$ confounding variable of unit $u$ and with $\mathbf{Z}$ a $P \times N$ matrix of random variables $Z_u^p$. 
As before, $Z^p$ will denote the $p^{th}$ row of $\mathbf{Z}$ and $Z_u$ its $u^{th}$ column. Our objective is to find pairs of units with minimum distance $\Delta$ as given in Eq. \eqref{eq:continuousDistance}. However, if the treatment and/or any of the confounding variables are noisy, the real distance cannot be calculated. Consequently, the applied matching method may result in poor matches. We attempt to improve the matching by including our knowledge about the uncertainty of the variables into the matching process. Suppose we have a notion of a distance $D(X_u, X_v)$ between random variables $X_u$, $X_v$ and a distance $\mathbf{D}(Z_u, Z_v)$ between random vectors $Z_u$ and $Z_v$. We need to find pairs of units $u$, $v$ that minimize

\begin{equation}
	\Delta(u, v) = \frac{\mathbf{D}(Z_u, Z_v)}{D(X_u, X_v)}
	\label{eq:probabilisticDistance}
\end{equation}

We need to define a suitable distance metric $D$ for our random variables. Commonly used distance metrics for distributions such as f-divergence metrics (e.g. Kullback-Leibler divergence) are not suitable in our case, since our objective is to estimate the probability that the values of two random variables $X_u$, $X_v$ are close (i.e. $Pr(|X_u - X_v|< \epsilon)$, where $\epsilon$ a small positive constant). Since our distance metric needs to measure also the proximity between the values of two random variables, we suggest a metric that is based on comparison of the quantiles of the examined variables. Let us denote by $q_{X_u}(k)$ the $k^{th}$ quantile of variable $X_u$, $k=1,2,...,K$. Then, we define $D(X_u, X_v)$ as follows:

% In order to better understand this, consider the three probability distributions depicted at Figure \ref{fig:exampleProbs}. In this example, the random variable $X_w$ is more likely than $X_v$ to have the same value with $X_u$ (i.e. $Pr(X_u = X_w) > Pr(X_u = X_v)$). However, the random variable $X_v$ is more likely to take a value that is 'close' to $X_u$ value (i.e. $Pr(|X_u - X_v|< \epsilon) > Pr(|X_u - X_w|< \epsilon)$). The Kullback - Leibler divergence for two variables $X_u$, $X_v$ is estimated as follows:
%
% \begin{equation*}
% 	D_{KL}(X_u||X_v) = \int Pr(X_u=x)\cdot \frac{log(Pr(X_u=x))}{log(Pr(X_v=x))} dx
% \end{equation*}
%
% \noindent
% In order to obtain a symmetric distance metric we estimate distance between $X_u$ and $X_v$ as $D_{KL}(X_u, X_v) = D_{KL}(X_u||X_v) + D_{KL}(X_v||X_u)$. For the examined distributions, we have $D_{KL}(X_u, X_v) = 4.94$ and $D_{KL}(X_u, X_w) = 3.69$, denoting that $w$ is a better match for $u$ than $v$, although $X_v$ is more likely to take values close to $X_u$ values.
%
%
% \begin{figure}
%   \centering
%   \includegraphics[width=0.45\textwidth]{/Users/ttsapeli/Dropbox/Birmingham/Birmingham/code/Matching-master/myCode/matlabCode/examplePlots.eps}
% \caption{Example Random Variables. }
% \label{fig:exampleProbs}
% \end{figure}

\begin{equation}
	D(X_u, X_v) = \frac{1}{K} \cdot \sqrt{\sum_k (q_{X_u}(k) - q_{X_v}(k))^2}
	\label{eq:randomVarDistance}
\end{equation}

If $X$ is not noisy, the quantile values will be the same and $D(X_u, X_v)$ reduces to the Euclidean distance of ${x}_u$ and ${x}_v$. % Considering again the example distributions of Figure \ref{fig:exampleProbs}, we have $\delta(X_u, X_v) = 0.22$ and $\delta(X_u, X_w) = 0.43$, denoting $v$ as better match for $u$.

\subsection{Probabilistic Genetic Matching}

Although several distance metrics can be used as the distance between random vectors $\mathbf{D}(Z_u, Z_v)$, in this work, we propose Probabilistic Genetic Matching (\emph{ProbGenMatch}), a modified version of the Genetic Matching distance metric. Denote by $\mathbb{D}_{Z_u, Z_v} = [D(Z_u^1, Z_v^1), D(Z_u^2, Z_v^2), ..., D(Z_u^P, Z_v^P)]^T$
the $P \times 1$ vector of distances $D(Z_u^p, Z_v^p)$ between the $P$ random variables $Z_u^p$, $Z_v^p$, $p=1,2,...,P$ (see Eq. \eqref{eq:randomVarDistance}). Then, we calculate $\mathbf{D}(Z_u, Z_v)$ by modifying the Genetic Matching distance of Eq. \eqref{eq:genMatch} as follows:

 \begin{equation}
 	\mathbf{D}(Z_u, Z_v) = \sqrt{\mathbb{D}_{Z_u, Z_v}^T \cdot (S^{-\frac{1}{2}})^T \cdot W \cdot S^{-\frac{1}{2}} \cdot \mathbb{D}_{Z_u, Z_v}}
 \end{equation}

The loss function used to select the optimal weight matrix $W$ also needs to be modified. We use the quantiles-based loss functions described in Section \ref{sec:matchingMethods}. In particular, for each pair of units $(u, v) \in G$ we define a random variable:

\begin{equation}
	{\cal A}^{p}_{u,v} = \frac{|Z^p_u - Z^p_u|}{|X_u - X_v|}
\end{equation}

We denote by $a^p_{u,v}(k)$ the $k^{th}$ quantile of ${\cal A}^{p}_{u,v}$. We also define the average $k$-th quantile for the $p^{th}$ confounding variable, $\bar{a}^p(k) = \frac{1}{|G|} \cdot \sum_{(u, v) \in G} a_{u,v}^p(k)$. Finally, we collect the average quantiles in the set $\Delta^p = \{\bar{a}^p(k)\}_{k=1}^K$ to be used in a quantile-based loss functions described
in Section \ref{sec:matchingMethods}.

\subsection{Implementation}

\emph{ProbGenMatch} has been implemented as an R package and it is based on the \emph{Matching} R package, an open source software which implements several matching methods. \emph{ProbGenMatch} takes as input the probability distributions for all the confounding variables and treatment variable for all the units of the study (along with other optional parameters) and returns the matched pairs according to the previously described framework. If all the variables of the study are observed without any measurement errors, then \emph{ProbGenMatch} is equivalent to the continuous Genetic Matching approach that we presented in Section \ref{sec:matchingContinuous}. If also the treatment variable is binary, \emph{ProbGenMatch} is equivalent to the Genetic Matching framework \cite{diamond2013genetic}. 

% \textbf{Many-to-one Matching.} The application of many-to-one matching or matching with replacement (discussed at section \ref{sec:MatchingParameters}) in scenarios with continuous treatments is not straightforward since units cannot be grouped to treated and control. Previously presented matching methods with continuous treatments are using one-to-one matching i.e. each unit can be used only one time. In our implementation, we offer to scientists the option to use each unit multiple times. In detain, scientists can decide about the maximum number of times $M$ that units are allowed to be used. Since this may result in some units being used multiple times while others not, we use frequency weights in order to eliminate the induced bias. Thus, for each matched pair $(u, v) \in G$ we assign a frequency weight:
%
% \begin{equation*}
% 	f_{(u,v)} = 0.5 \cdot \Big(\frac{1}{n_u} + \frac{1}{n_u} \Big)
% \end{equation*}
% \noindent
% where $n_u$, $n_v$ the number of times the units $u$, $v$ have been used respectively.

\textbf{Caliper Distance.} Matching with caliper distance has been previously proposed as a way to impose restrictions on the maximum allowed dissimilarity between the matched units \cite{austin2011introduction}. A caliper distance is simply a threshold that defines the maximum allowed difference of two units on their confounding variable values.
In our implementation we also support matching with caliper distance as an optional parameter. For stochastic confounding variables, a probability threshold $T_{prob}$ should be provided along with the caliper distance. This probability threshold allows the matching of two units only if the probability to have a larger difference than the caliper distance on their confounding variable values is smaller than $T_{prob}$. Our implementation also allows users to specify a threshold on the minimum difference between the treatment values of two matched units. For stochastic treatment variables, a probability threshold should be provided along with the minimum treatment difference threshold.

%
%
% \textbf{Computational Cost.} \emph{ProbGenMatch} requires more computational resources than traditional genetic matching. In detail, the cost of estimating the distance between two confounder vectors, as described at eq. \eqref{eq:genMatch} is $O(P)$ and the cost of estimating the distances between all units pairs is $O(P \cdot N^2)$. In contrast, \emph{ProbGenMatch} requires $O(K)$ in order to estimate the distance between two random variables, as described by eq. \eqref{eq:randomVarDistance}, $O(K\cdot P)$ for the estimation of the distance between two confounder vectors and $O(K \cdot P \cdot N^2)$ for the estimation of the distances between all units pairs.

\section{Evaluation}
\label{sec:evaluation}
We evaluate the proposed probabilistic matching framework on two synthetic and one real dataset. All the examined scenarios include one unobserved variable $L$ along with an observed noisy version $\tilde{L}$. We use as baselines for our evaluation:

\begin{enumerate}
	\item the traditional Genetic Matching (\emph{GenMatch}) approach which treats $\tilde{L}$ as the \emph{true} variable. 
	\item the \emph{optimal} Genetic Matching (\emph{OptGenMatch}), where we assume that $L$ is observed without any noise. The performance of Genetic Matching under this optimal scenario serves as an upper bound to the performance of our method. The results obtained by \emph{OptGenMatch} will be considered as ground-truth. 
\end{enumerate}	

We use the synthetic datasets to evaluate the performance of the proposed framework on different noise levels. We also examine the sensitivity of our approach to the selected caliper distance. Finally, we apply our method on the social media dataset described in \cite{MicroBlogPCU} in order to test whether text messages containing URLs tend to be reposted more often.

\textbf{Evaluation Criteria.} The objective of a matching method is to match units with small dissimilarity on their confounding variables values and large difference on their treatment values. We evaluate the performance of \emph{ProbGenMatch} on these metrics in comparison with \emph{GenMatch} and \emph{OptGenMatch}. Additionally, we investigate whether the amount of the remaining bias due to purely matched pairs is sufficiently large to influence the validity of the causal inference by comparing our conclusions with the corresponding results of \emph{OptGenMatch}.

\subsection{Synthetic Dataset}

We consider a binary variable $L$ describing the class of objects represented by M-dimensional vectors of real numbers. We consider two types of vectors. The first type corresponds to positive examples (i.e. vectors that belong to class $L=1$). The data in each of the $M$ dimensions of the first vector type are generated by a Gaussian process with mean value $1$ and standard deviation $\sigma_1$. The second type of vectors corresponds to negative examples (i.e. $L=0$) and their values in each dimension are generated by a random Gaussian process with mean $-1$ and standard deviation $\sigma_2$. We train a Support Vector Machine classifier on this synthetic dataset and afterwords we use the classifier on unseen synthetic data (generated with the same procedure) in order to learn a label $\tilde{L}$ for each vector. Afterwards, we map the SVM outputs into probabilities by applying the process described in \cite{platt1999probabilistic}. For each vector $v$, the probability distribution of random variable $L_v$ corresponds to the output of this mapping procedure. 

In our test case, we consider two-dimensional vectors (i.e. $M=2$) and we set $\sigma_1 = 1$. We test the performance of our matching framework with different noise levels on the observed variable $\tilde{L}$ by increasing $\sigma_2$ from $1$ to $2$ with step $0.2$. By increasing the variance of the second vector type, we make our vectors less separable and consequently, the resulted classes $\tilde{L}$ are less accurate. 

\subsubsection{Unobserved Treatment Variable}
\label{sec:bayesTr}
\begin{figure}
  \centering
  \includegraphics[width=0.5\textwidth]{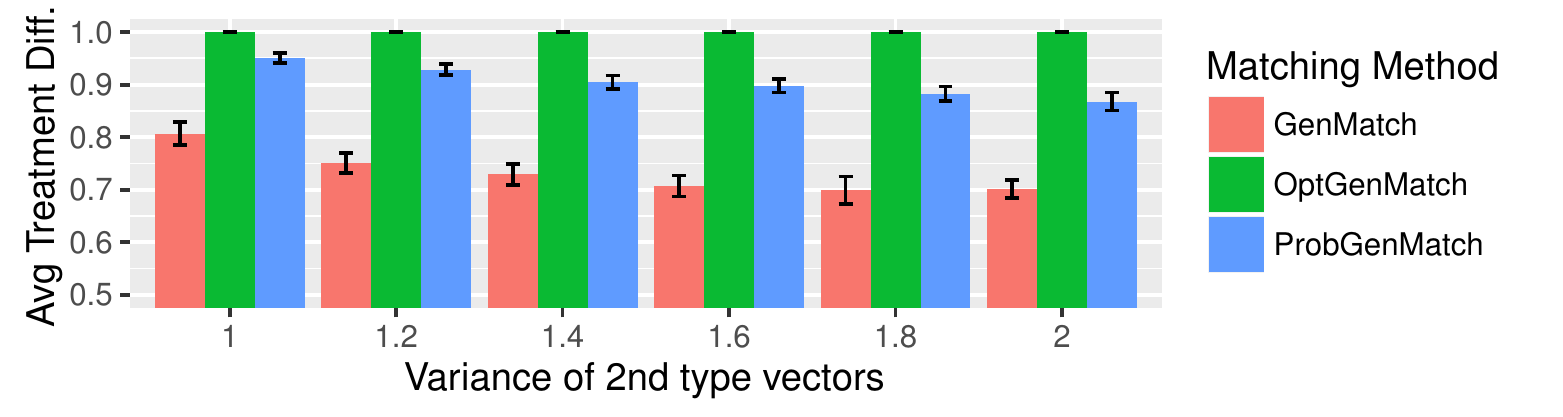}
\caption{Average treatment difference between the matched units.}
\label{fig:naiveBayesAvgTrDiff}
\end{figure}

In the first case, we consider $L$ as the treatment variable. We generate two confounding variables $Z^1 = \alpha_1 \cdot L + e_1$ and $Z^2 = \alpha_2 \cdot L + e_2$, where $e_1$ and $e_2$ random Gaussian noise with mean $0$ and variance $1$ and $2$ respectively for $Z^1$ and $Z^2$ and $\alpha_1, \alpha_2$ are model coefficients. In the following results, we do not use a caliper distance for the confounding variables. We set the minimum allowed distance between the treatments of matched units equal to $0.1$ and the maximum allowed probability that the matched units have a treatment difference larger than $0.1$ equal to $0.25$. We repeat our study for 10 randomly selected sets of model coefficients ($\alpha$s). All model coefficients are randomly generated from a uniform distribution on $[0, 1]$. For each one of the 10 sets of model coefficients we repeat each study for 100 different noise realizations. In Fig. \ref{fig:naiveBayesAvgTrDiff} we present the average treatment difference between the matched units for the three examined matching algorithms along with the 95\% confidence intervals. The \emph{OptGenMatch} method always avoids matching units with the same treatment value. Thus, given that in this scenario we consider binary treatments, the average treatment difference is always equal to $1$. According to our results, the performance of both \emph{GenMatch} and \emph{ProbGenMatch} declines for higher noise levels (i.e. larger $\sigma_2$). However, \emph{ProbGenMatch} significantly outperforms \emph{GenMatch} by avoiding matching units with the same treatment for more than 88\% of the matched pairs for all examined noise levels.

\begin{figure}
  \centering
  \includegraphics[width=0.5\textwidth]{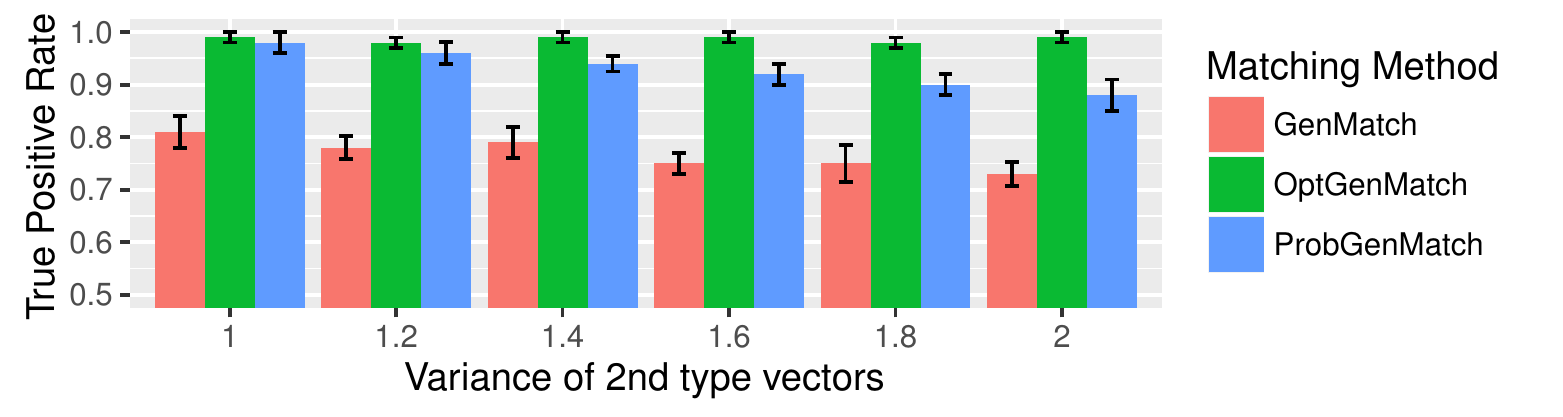}
\caption{Percentage of true positive causality conclusions.}
\label{fig:naiveBayesCausalityTr}
\end{figure}

When the resulted group of matched units contains pairs with the same treatment level, the impact of the examined treatment on the outcome variable cannot be reliably assessed by comparing the matched units on their outcome values. We demonstrate this by generating the following outcome variable:

\begin{equation} 
Y  = \beta_0 \cdot L +  \beta_1 \cdot Z^1 + \beta_2 \cdot Z^2 + n_u + e_n
\label{eq:outcomeVariable}
\end{equation}
where $\beta_0, \beta_1, \beta_2$ are model coefficients, $n_u$ is a uniform random variable on $[0, 4]$ and $e_n$ is gaussian noise with mean $0$ and variance $1$. All $\beta$ coefficients are randomly generated from a uniform distribution on $(0, 1]$. For non-zero $\beta_0$, the treatment variable $L$ has a causal impact on $Y$. We apply a Wilcoxon non-parametric test in order to examine whether the average treatment effect, (Eq. \eqref{eq:avgTrEffect})  is significantly different than zero. When the performance of \emph{OptGenMatch} is examined, we use as treatment (i.e. the variable $X$ of Eq. \eqref{eq:avgTrEffect}) the binary variable $L$, while for \emph{GenMatch} and \emph{ProbGenMatch} we use the noisy variable $\tilde{L}$. We repeat our study for $10$ different sets of model coefficients and $100$ realizations of $n_u$, $e_n$, for all the groups of matched units resulted after applying the three examined methods, as it was previously described. In Fig. \ref{fig:naiveBayesCausalityTr} we depict the average percentage of times that the null hypothesis of the Wilcoxon test (i.e. that the average treatment effect is equal to zero) was rejected with p-value equal to $0.05$. \emph{OptGenMatch} successfully detects the causal impact of $L$ on $Y$ in most cases, while \emph{ProbGenMatch} significantly outperforms \emph{GenMatch}.  
	
\subsubsection{Latent Confounding Variable}
\label{sec:bayesConf}

In the second case, $L$ corresponds to a binary confounding variable. In detail, we consider a continuous treatment variable $X$ which follows a uniform distribution on $[0, 1]$. Our binary confounding variable $L$ follows a binomial distribution with success probabilities given by the vector of probabilities $P_S = 1 / (1+e^{-t})$, where $t = \alpha_0 + \alpha_1 \cdot X$. We also create a confounding variable $Z^1 = \alpha_1 \cdot X + e_1$. We evaluate the performance of the three examined matching approaches by generating different realizations of the model coefficients and noise $e_1$, as it was previously described in Section \ref{sec:bayesTr}. We assess the remaining bias due to imperfect matches by calculating the standardized difference in means for each confounding variable \cite{MatchingSurvey}. In detail, for the binary confounding variable $L$, we consider the values $\{\frac{l_u}{x_u-x_v}: (u, v) \in G\}$ as realizations of a random variable $C_U$ and the values $\{\frac{l_v}{x_u-x_v}: (u, v) \in G\}$ as realizations of a random variable $C_V$. Then the standardized difference in means for the confounding variable $L$ is estimated as:

\begin{equation}
	\frac{|\bar{C}_U - \bar{C}_V|}{\sqrt{(\sigma_{U}^2 + \sigma_{_V}^2)/2}}
	\label{eq:smd}
\end{equation}
where $\bar{C}_U$, $\bar{C}_V$  the mean values of $C_U$, $C_V$ respectively and $\sigma_U^2$, $\sigma_V^2$ their variances. The same process is followed for the estimation of the standardized difference in means for $Z^1$. 

% For the continuous confounding variable $Z^2$ the remaining bias is estimated as:
%
% \begin{equation}
% 	\frac{\mu_{Z^2,U} - \mu_{Z^2,V}}{\sqrt{(\sigma_{Z^2,U}^2 + \sigma_{Z^2,V}^2)/2}}
% \end{equation}
% where $\mu_{Z^2,U}$, $\sigma_{Z^2,U}^2$, $\mu_{Z^2,V}$ and $\sigma_{Z^2,V}^2$ the mean value and the variance of the variable $Z^2$ for the units that belong to sets $U$ and $V$ respectively. For the binary confounding variable $L$ the remaining variance is estimated as follows \cite{}:
%
% \begin{equation}
% 	\frac{\mu_{L,U} - \mu_{L,V}}{\sqrt{(\mu_{L,U}\cdot(1-\mu_{L,U})+\mu_{L,V}\cdot(1-\mu_{L,V}))/2}}
% \end{equation}
 
\begin{figure}
  \centering
  \includegraphics[width=0.5\textwidth]{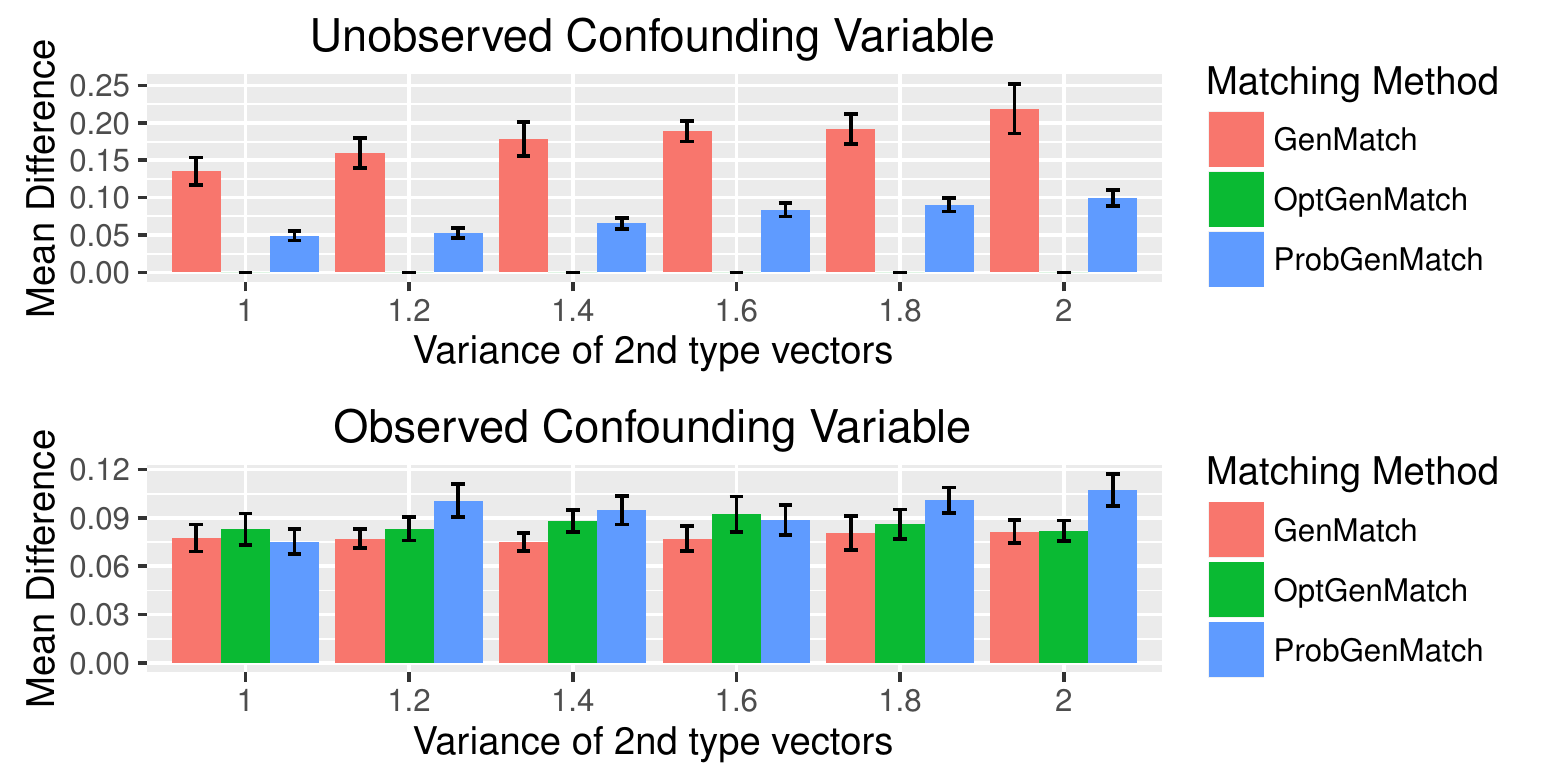}
\caption{Remaining bias for the two confounding variables.}
\label{fig:naiveBayesBiasconf}
\end{figure}

In Fig. \ref{fig:naiveBayesBiasconf} we present the standardized difference in means for the two confounding variables (i.e. the binary variable $L$ on the top and the continuous $Z^1$ on the bottom). \emph{OptGenMatch} always matches units with the same value on $L$ and therefore, there is zero bias. The proposed \emph{ProbGenMatch} method achieves also low bias, smaller than $0.1$ for all the noise levels and significantly outperforms \emph{GenMatch}. Finally, all methods achieve similar performance on the continuous confounding variable $Z^1$, which is considered to be observed without any noise, although the performance of \emph{ProbGenMatch} is slightly worse for large noise levels. 

\begin{figure}
  \centering
  \includegraphics[width=0.5\textwidth]{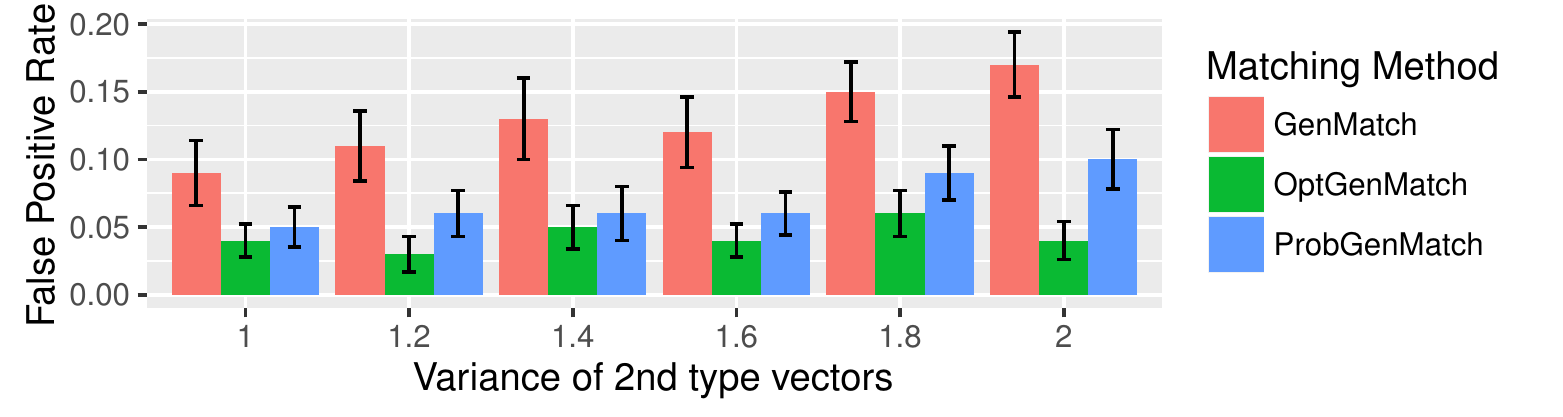}
\caption{Percentage of false positive causality conclusions.}
\label{fig:naiveBayesCausalityConf}
\end{figure}

Failing to sufficiently eliminate the bias induced by confounding variables may result in false positive causality conclusions. We demonstrate this by considering again the outcome variable of Eq. \eqref{eq:outcomeVariable}. This time we set $\beta_1 = 0$, thus, there is no causal impact of $Z^1$ on $Y$. In Fig. \ref{fig:naiveBayesCausalityConf} we present the rate of the false positive causality conclusions (i.e. the average percentage of times that the null hypothesis of the Wilcoxon test was not rejected) along with the 95\% confidence interval. \emph{ProbGenMatch} achieves up to 8\% lower false positive rate than \emph{GenMatch}. 

\subsection{Location-based Synthetic Dataset}

In this scenario, our latent variable $L$ represents the daily time that the participants of a study spend in entertainment venues such as pubs, restaurants, bars etc. We assume a study based on stmartphones sensor data, where participants do not report their location; instead location, along with the underlying context (i.e. work, home, restaurant etc.) is inferred from other raw sensor data. Several methods for automatic location label inference have been proposed \cite{chon2012automatically, zhou2007discovering}. However, the real location context cannot be inferred accurately. Location context could be very important for studies examining the impact of social behavior or daily activities (e.g. exercising socializing etc.) on well-being indicators such as stress level \cite{tsapeli2015investigating} or eating disorders \cite{madan2010social}. 

We synthetically generate a location dataset based on the description of the real dataset presented in \cite{chon2012automatically}. In \cite{chon2012automatically}, authors gather several sensor data along with ground truth labels for the locations of 36 participants and they apply a method for automatic location label inference. In order to generate our dataset, we define a variable $P$ denoting a location label. As described in \cite{chon2012automatically}, we consider 7 location labels: \emph{home}, \emph{work}, \emph{college}, \emph{entertainment}, \emph{food}, \emph{shops} and \emph{other}. $P_u(t)$ denotes the location of a participant at day $u$ and time $t$ and it is sampled on an hourly basis. We use the mobility patterns and location statistics described in \cite{chon2012automatically} in order to randomly generate the variable $P_u(t)$. We also define a variable $E_u(t)$ as follows:

\begin{equation}
	E_u(t) = \begin{cases}
	1 &\text{, $P_u(t) = $\emph{entertainment}}\\
	0 &\text{, otherwise}
	\end{cases}
\end{equation}

Finally, we create a variable $L$, with values $l_u = \sum E_u(t)$ for each day $u$. However, in a real study, where participants would probably be unwilling to continuously provide labels for their location data, $L$ would be a latent variable. We generate the discrete variable $\tilde{P}(t)$ denoting the inferred location label based on the method described in \cite{chon2012automatically} by utilizing the confusion matrix (Table 3 of \cite{chon2012automatically}) which presents the performance of the proposed location inference method. According to this matrix, only 41\% of the places with resulted label \emph{entertainment} are correctly labeled while the rest 59\% of the places actually correspond to \emph{college} (4\%), \emph{work} (4\%), \emph{shops} (4\%), \emph{food} (33\%) and \emph{others} (9\%). We create a noisy variable $\tilde{P}(t)$ by randomly inserting bias on $P(t)$ based on these results. Then we define $\tilde{E}_u(t)$ as:

 \begin{equation}
 	\tilde{E}_u(t) = \begin{cases}
 	1 &\text{, $\tilde{P}_u(t) = $\emph{entertainment}}\\
 	0 &\text{, otherwise}
 	\end{cases}
 \end{equation}
We also create $\tilde{L}_u$ with values $\tilde{l}_u = \sum \tilde{E}_u(t)$ for each day $u$. Finally, based on the performance of the location inference method, we create a random variable $L_u$ with probability distribution $Pr(L_u|\tilde{P}_u(1), \tilde{P}_u(2), ... \tilde{P}_u(24))$. We normalize $L$, $\tilde{L}$ to $[0, 1]$. 

% \subsubsection{Latent Treatment Variable}

\begin{figure}
  \centering
  \includegraphics[width=0.5\textwidth]{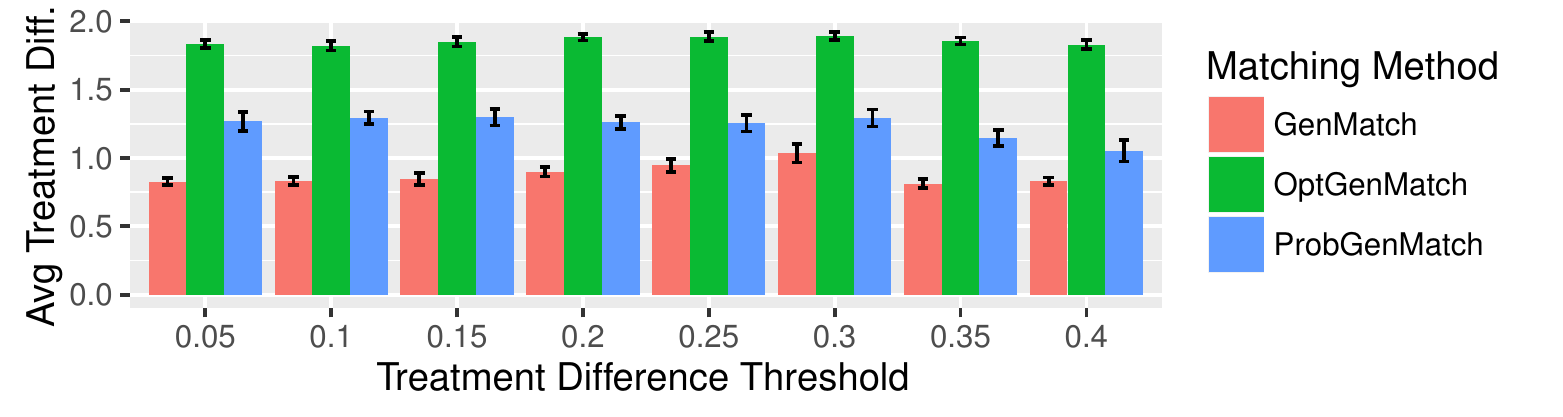}
\caption{Average Treatment Difference between the matched units.}
\label{fig:locationTrDiff}
\end{figure}

We use $L$ as the unobserved treatment variable and we generate the confounding variables $Z^1$, $Z^2$ as it is described in Section \ref{sec:bayesTr}. In this scenario, we examine the impact of the allowed minimum treatment difference on the three examined matching methods. In detail, let us denote with $T_{min}$ the minimum allowed treatment distance. We vary $T_{min}$ from $0.05$ to $0.4$ with $0.05$ step. For  \emph{ProbGenMatch} we set the maximum allowed probability that the treatment difference is smaller than $T_{min}$ equal to $0.25$. In Fig. \ref{fig:locationTrDiff} we present the average treatment difference between the matched treated and control groups achieved by the three examined matching algorithms. According to our results, there is not significant impact of the treatment difference threshold on the average treatment difference when the \emph{OptGenMatch} method is applied. There is an improvement on the performance of \emph{GenMatch} for the threshold values 0.2 to 0.3, however its performance is decreased for larger than 0.3 thresholds. Since the threshold is applied on the observed noisy variable $\tilde{L}$ and not on $L$, large threshold values may prevent the matching of units which are actually good matches. \emph{ProbGenMatch} is not strongly influenced by the treatment difference threshold, however, its performance is also decreasing for large threshold values. 

\begin{figure}
  \centering
  \includegraphics[width=0.5\textwidth]{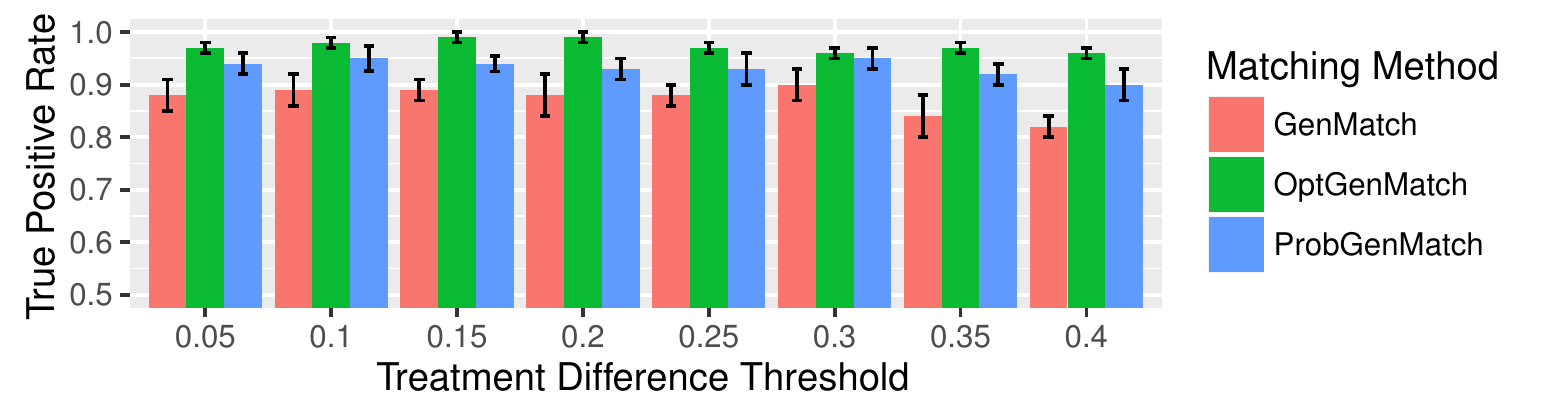}
\caption{Percentage of True Positive Causality Conclusions. }
\label{fig:locationTrTPCausality}
\end{figure}

Finally, we generate again an outcome variable as described in Eq. \eqref{eq:outcomeVariable} in order to examine the influence  that the resulted matching may have on a causality study. We examine the rate of true positive causality conclusions for the three examined methods by repeating the process described in Section \ref{sec:bayesTr} and we present our results in Fig. \ref{fig:locationTrTPCausality}. \emph{ProbGenMatch} achieves a higher rate of true positive conclusions compared to \emph{GenMatch}, however their difference is less significant compared to the binary treatments case examined in Section \ref{sec:bayesTr}. This is reasonable considering that for binary noisy treatments matching will result more often on pairs with the same treatment value; thus, the treatment effect will be weaker. 

\subsection{Social Media Dataset}

In this section, we evaluate our method on a real dataset. We use the microblogPCU dataset, which is available in the UCI Machine Learning repository \cite{MicroBlogPCU}, in order to examine whether the number of URLs included in microblog messages influences the number of times that these messages are re-posted. MicroblogPCU dataset has been collected from the sina Weibo microblog and contains information about the profiles of 782 users, their social network and their microblog activity. It also contains ground-truth binary labels indicating whether a user is a \emph{spammer} or not for 111 users. 

We use the ratio of messages with URLs as the treatment variable of our study and the number of re-posts as the outcome variable. Spammers tend to use more URLs in their messages and spammers messages are re-posted less often. Thus the spammer binary indicator should be used as a confounding variable. We also use as confounding variables other indicators that correlate both with the treatment and the outcome variables. In detail, we found that the number of posts, the class level of the user account (this is an indicator assigned by Weibo) and the number of followers correlate both with our treatment and outcome variables. 

We assume that the binary spammer indicator is unknown and it needs to be inferred from the data. We apply the method described in \cite{benevenuto2010detecting} in order to classify the users to spammers and non-spammers. We extract attributes from the text content and users profiles as described in \cite{benevenuto2010detecting}. We use all the attributes of \cite{benevenuto2010detecting} apart from the number of times a user replied to a message or received a reply and whether a message is a \emph{reply} message, since this information is not provided in this dataset. Also, instead of the user account age, we use the user account class. The interested reader should refer to \cite{benevenuto2010detecting} for a complete list of all the extracted features.  Afterwards, we apply the chi-squared feature selection method in order to find the most important attributes. Six attributes were selected, namely: 1) the fraction of tweets with URLs, 2) the user account class, 3) the average number of URLs per message, 4) the number of followees, 5) the average number of hashtags per message and 6) the average number of re-posts. Following the procedure of \cite{benevenuto2010detecting}, we use weka \cite{holmes1994weka} to train a support vector machine classifier. We used only the data for the 111 users for which a ground-truth label is available. We used 50\% of the dataset for training and the rest for testing. 76\% of the spammers and 82\% of the non-spammers are correctly classified. Our classifier is more successful on recognizing the spammers and less on recognizing the non-spammers compared to \cite{benevenuto2010detecting}. This difference can be attributed to the differences between the dataset characteristics of the two studies.

We define as $L$ the ground-truth binary label indicating whether a user is spammer and $\tilde{L}$ the inferred label based on the above-mentioned process. We also create a random variable $L_u$ for each user $u$ and we obtain a probability distribution for each $L_u$ by mapping the SVM outputs into probabilities. We match the users based on their confounding variables values by applying the three examined approaches. The results obtained by \emph{OptGenMatch} serve as the ground-truth.
Finally, we use the Wilcoxon test to examine whether the mean value of the outcome variable for the treated units significantly differs from the mean outcome value of the control units. In Table \ref{tab:weiboRes} we present the mean difference (see eq. \eqref{eq:smd}) achieved for the binary confounding variable $L$ with the 3 examined methods. We also present the p-values of the Wilcoxon test under the null hypothesis that the treatment variable has no effect. Both \emph{OptGenMatch} and \emph{ProbGenMatch} reject the null hypothesis with p-value smaller than 0.05. However, when the treatment and control pairs are created by applying the \emph{GenMatch} method, the remaining bias on the binary indicator $L$ is large and results in the false conclusion that there is no significant impact of the number of URL's included in text messages to the number of re-posts.  

\begin{table}[]
\centering
\begin{tabular}{|c|c|c|}
	\hline
 	{}& \textbf{Balance on L} & \textbf{Wilcoxon test p-value}\\
  \hline
 \textbf{OptGenMatch}& 0.014 & 0.005\\
  \hline
   \textbf{GenMatch}& 0.36 & 0.15\\
 \hline
   \textbf{ProbGenMatch}& 0.15 & 0.041\\
 \hline

\end{tabular}
\caption{Causality Study Results}
\label{tab:weiboRes}
\end{table}

\section{Related Work}

Causal inference when important variables are missing is mainly based on structural equation models (SEMs) with latent variables \cite{fornell1981structural, bentler1980multivariate}. SEMs include two components: the structural model part describing the causal relationships between the predictors and the outcome variables and the measurement model describing the relationships of the latent variables with other observed variables. Model selection is based on theoretical assumptions about data structure. After the model is specified it should be assessed whether the data fit on it by using statistical tests or fit indexes. Graphical causal models with latent variables have also been used \cite{meganck2007causal}. Our approach examines the problem of unobserved variables on a different causality framework (i.e. the matching framework), thus it should be considered complementary rather than competitive to the existing methods. SEMs are usually based on assumptions about the model-class of the data, while matching methods require fewer assumptions. Moreover, assessing the model bias with matching-based methods is straightforward by using balance diagnostics tests while, goodness-of-fit tests used with structural equation models cannot assess whether systematic differences between units have been eliminated \cite{hirano2004propensity}. Thus, causal inference based on matching could be preferable in some cases. To the best of our knowledge, our method is the first that considers the problem of unobserved factors on matching methods.  
 
\section{Conclusions}
We propose \emph{probabilistic matching}, a novel approach for causal inference when one or more key variables are unobserved or noisy. Our method is based on the assumption that probability distributions describing the values of the unobserved variables are known or can be approximated. We define a distance metric, based on Genetic Matching distance, that measures the dissimilarity between units by examining the difference on the quantiles of the stochastic variables of the study. We evaluate our method both on simulated and real datasets. We demonstrate that when the treatment variable is noisy, traditional matching methods may result in matching pairs with the same or similar treatment values. This weakens the observed effect of the treatment on the outcome variable and consequently, may result in missing true causal links. Similarly, when there is noise on the confounding variables, existing matching methods fail to sufficiently reduce the confounding bias, which often results in false causality conclusions. We show that our approach is able to find better matches and, consequently, achieves more accurate causal conclusions.

% \bibliography{myBib}
% \bibliographystyle{unsrt}

\end{document}